\documentclass[aps,prl,reprint,floatfix,superscriptaddress]{revtex4-1}

\usepackage{graphicx}
\usepackage{hyperref}
\usepackage{amsmath}
\usepackage{tabularx}

\begin{document}


\title{Electric-field-induced energy tuning of on-demand entangled-photon emission from self-assembled quantum dots}

\author{Jiaxiang Zhang}
\email[]{jiaxiang.zhang2012@gmail.com}
\affiliation{Institute for Integrative Nanosciences, IFW Dresden, Helmholtzstra$\beta$e 20, 01069 Dresden, Germany}
\author{Eugenio Zallo}
\affiliation{Institute for Integrative Nanosciences, IFW Dresden, Helmholtzstra$\beta$e 20, 01069 Dresden, Germany}
\affiliation{Present address: Paul-Drude-Institut f\"{u}r Festk\"{o}rperelektronik Hausvogteiplatz 5-7, 10117 Berlin, Germany}
\author{Bianca H\"{o}fer}
\author{Yan Chen}
\author{Robert Keil}
\author{Michael Zopf}
\author{Stefan B\"{o}ttner}
\author{Fei Ding}
\affiliation{Institute for Integrative Nanosciences, IFW Dresden, Helmholtzstra$\beta$e 20, 01069 Dresden, Germany}
\author{Oliver G. Schmidt}
\affiliation{Institute for Integrative Nanosciences, IFW Dresden, Helmholtzstra$\beta$e 20, 01069 Dresden, Germany}
\affiliation{Material Systems for Nanoelectronics, TU Chemnitz, Reichenhainerstra$\beta$e 70, 09107 Chemnitz, Germany}


\date{\today}

\begin{abstract}
The scalability of quantum dot based non-classical light sources relies on the control over their dissimilar emission energies.~Electric fields offer a promising route to tune the quantum dot emission energy through the quantum-confined Stark effect.~However, electric fields have been mostly used for tuning the energy of single-photon emission from quantum dots, while electrical control over the energy of entangled-photon emission,~which is crucial for building a solid-state quantum repeater using indistinguishable entangled photons, has not been realized yet.~Here, we present a method to achieve electrical control over the energy of entangled-photon emission from quantum dots.~The device consists of an electrically-tunable quantum diode integrated onto a piezoactuator.~We find that, through application of a vertical electric field, the critical uniaxial stress used to eliminate the fine-structure-splitting of quantum dots can be linearly tuned.~This allows realization of a triggered source of energy-tunable entangled-photon emission, an important step towards a solid-state quantum repeater application.
\end{abstract}


\maketitle

On-demand sources of entangled-photon pairs play a central role in quantum information science, and are essential for implementing photonic quantum technologies such as quantum key distribution \cite{Gisin02}, quantum teleportation \cite{Bouwmeester97} and quantum computation \cite{Knill01,Kok07}.~To build a scalable quantum network, entanglement distribution  between remote quantum nodes is required by extending entangled-photon pairs using indistinguishable photons in Bell state measurements \cite{Kimble08}.~This process, called entanglement swapping, is the heart of a quantum repeater proposed to increase the distance of quantum communication.~So far, the most widely used entangled-photon sources for realizing entanglement swapping are based on the spontaneous parametric down-conversion process in nonlinear optical media \cite{Riedmatten04,Pan98,Riedmatten05}, but they produce a probabilistic number of entangled-photon pairs per excitation cycle.~This probabilistic nature would cause errors in quantum algorithm protocols and thus fundamentally limits successful use of such sources in deterministic quantum applications \cite{Scarani04}. 
  
~Semiconductor quantum dots (QDs), allowing generation of non-classical photons with near-unity probability,~are among the most promising entangled-photon sources.~They offer many key features towards experimental realization of entanglement swapping, including ultra-bright \cite{Douse10,Versteegh14}, highly indistinguishable polarization entangled-photon emission \cite{Muller14} and easy integration with a diode structure to realize electrical excitation \cite{Benson00,Salter10,Zhang15}.~In practice, however, self-assembled QDs suffer from a random growth process, which leads to mesoscopic structure, strain and composition anisotropies.~As a result, realization of entanglement swapping with QDs encounters two main challenges: elimination of the energetic splitting between the two bright exciton (X) states known as the fine structure splitting (FSS) \cite{Gammon96,Bayer02}, and energy tuning of emitted photons simultaneously. Despite the great success of eliminating the FSS ($s$) by employing a variety of post-growth tuning techniques such as in-plane magnetic field \cite{Stevenson05}, anisotropic strain fields \cite{Zhang15} and optical stack effect \cite{Muller09}, the reported QDs entangled-photon sources are usually restricted to different emission energies due to the lack of simultaneous control over the energy of entangled-photon emission.~This hurdle becomes the main obstacle impeding Bell state measurements between entangled photons emitted by remote QDs. Therefore, the real potential of QDs for entanglement swapping can be harnessed only when a tight control over the energy of entangled-photon emission is achieved. 

It has been reported recently that energy-tunable entangled-photon emission from QDs can be achieved by applying two or three independent strain fields using microstructured piezoactuators \cite{Trotta16,Chen16}.~However, on-demand generation of energy-tunable entangled-photon pairs has not been demonstrated.~Indeed, realization of a QD source that emits no more than one entangled photon pair per excitation cycle has significant advantages in practical applications as it allows for a non-postselective two-photon interference at a beam splitter between two remote QDs \cite{Flagg10,Gold2014,Giesz15}.~Moreover, these microstructured piezoactuators consisting of six or four strain legs require sophisticated design and microfabrication techniques such as focused ion beam writing and femtosecond pulsed laser ablation, which renders practical applications more challenging.~Compared to the strain-field-induced energy tuning, applying electric fields to control the energy of the QD emission is practically favorable.~Considerable benefits of electric-field-induced energy tuning are the high stability and ultrafast response \cite{Patel2010} without hysteresis and creep that unavoidably appears for the strain-field-induced energy tuning \cite{Zander09,Trotta12,Rastelli12}.~In this sense, it is highly desirable to achieve electric-field-induced energy tuning of on-demand entangled-photon emission from QDs.  
     
\begin{figure}[h!]
	\centering
	\includegraphics[scale=1.0]{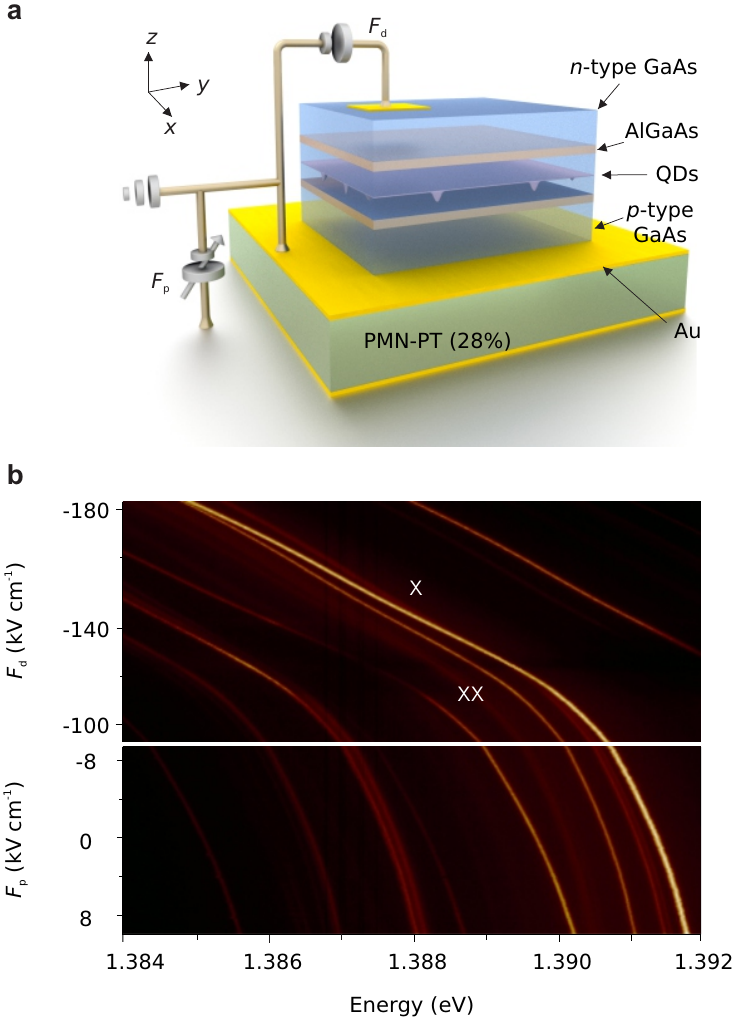}
	\caption{\textbf{$|$ Schematic illustration of the device and energy shift of the excitonic transitions induced by electric and strain fields independently}.~\textbf{a}, Sketch of the thin QDs-containing \textit{n-i-p} diode heterostructure.~\textbf{b}, Representative energy shift of the photoluminescence from a single QD as the electric fields $F_\text{p}$ and $F_\text{d}$ are applied to the PMN-PT actuator and the diode, respectively.}
\end{figure}

In this letter, we present a new approach to realize electric-field-induced energy tuning of on-demand entangled-photon emission from QDs.~This is achieved by embedding QDs inside an electrically tunable \textit{n-i-p} diode nanomembrane integrated onto a piezoactuator, which is capable of exerting uniform uniaxial stress to the entire QDs nanomembrane.~We theoretically and experimentally reveal that the magnitude of the uniaxial stress required to cancel the FSS of QDs can be linearly tuned by a vertical electric field through the quantum-confined Stark effect (QCSE).~Accordingly, triggered entangled-photon emission with a broad energy tuning range is realized.~High entanglement-fidelities at different photon energies are obtained in response to the vertical electric field and the external uniaxial stress.~The device demonstrated in this work represents a major step towards the experimental realization of entanglement swapping in semiconductor QD systems.
       
\begin{figure*}[t!]
	\centering
	\includegraphics[scale=0.9]{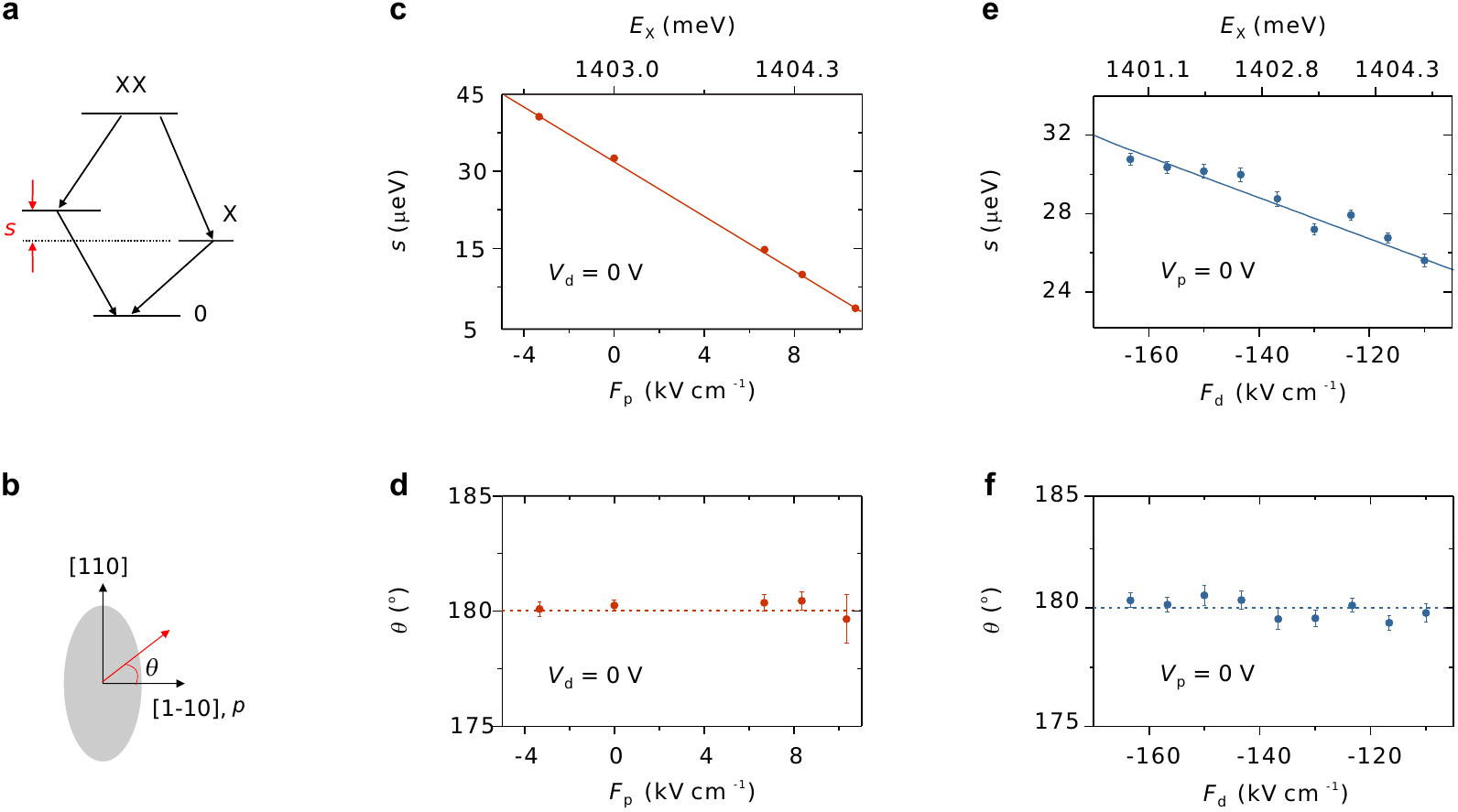}
	\caption{\textbf{$|$ Tuning of $s$ and $\theta$ through independent application of $F_p$ and $F_\text{d}$.} \textbf{a}, Schematic of the biexciton cascade emission in a single QD through which polarization-entangled two-photon can be generated. The energy separation between the two bright exciton states is referred to as the FSS. \textbf{b}, Schematic illustration of $\theta$ with respect to [1-10] crystal axis of the GaAs, which is precisely aligned along the uniaxial axis of $p$. \textbf{c} - \textbf{d}, Behaviors of $s$ and $\theta$ as a function of $F_\text{p}$ at $V_\text{d}$ = 0 V (corresponding to the diode built-in potential $F_\text{d}$ = -133 kV cm$^{-1}$). \textbf{e} - \textbf{f}, Behaviors of $s$ and $\theta$ as a function of $F_\text{d}$ at $V_\text{p}$ = 0 V. }
\end{figure*}

Fig.~1a shows a sketch of the device, which consists of a \textit{n-i-p} diode nanomembrane integrated onto a 300 $\mu$m-thick piezoelectric actuator, [Pb(Mg$_\text{1/3}$Nb$_\text{2/3}$)O$_3$]$_\text{0.72}$-[PbTiO$_3$]$_\text{0.28}$ (PMN-PT). The diode nanomembrane contains a layer of low density In(Ga)As QDs in the middle of the 150 nm-thick intrinsic GaAs/Al$_\text{0.4}$Ga$_\text{0.6}$As quantum well.~The quantum well is used to increase the barrier height for photo-excited charge carriers, which enables QD transition energies to be shifted over a broad spectral range utilizing the QCSE \cite{Bennett10}. Electrical contacts are subsequently made in such a way that electric fields can be independently applied to the diode and the PMN-PT actuator \cite{Zhang13,Trotta12,Trotta121,Trotta14,Zhang16}.~Although early works used a similar device, it was constrained to only eliminate the FSS of QDs by using the combination of an electric field and an anisotropic biaxial strain field \cite{Trotta121,Trotta14}. Extending this work to achieve energy tuning of entangled-photon emission is not straightforward as the anisotropic biaxial strain field cannot cancel the FSS \textit{solely} \cite{Plumhof11}.~Here, instead, we use a piezoelectric actuator with pseudo-cubic cut directions [100], [0-11] and [011] (denoted by \textit{x}, \textit{y}, \textit{z} axis respectively), and approximately uniaxial stress, $p$, along the \textit{y} axis can be expected in the QD nanomembrane (see Supplementary Note 1).~As demonstrated in the following, by using the combination of such uniaxial stress and the second vertical electric field tuning mechanism, on-demand energy-tunable entangled-photon emission can be realized.

When a reverse bias $V_\text{d}$ (electric field $F_\text{d}$) is applied to the diode and a voltage $V_\text{p}$ (electric field $F_\text{p}$) is applied to the PMN-PT actuator, energy-tunable photoluminescence (PL) from QDs can be produced as shown in Fig. 1b.~According to the power and polarization-resolved measurements, the brightest PL lines are assigned to the exciton (X) and biexciton (XX) photon emission from a single QD.~As the magnitudes of the electric fields are varied,~a total energy shift of about 7 meV for the transition energies of QDs is obtained. The energy shift induced by $F_\text{d}$ is due to the QCSE, while the strain-induced energy shift with $F_\text{p}$ is ascribed to the change of the volumetric strain $\varepsilon_\text{tot}$ at the QD position \cite{Zhang15}. In addition to this significant QD energy tunability, we now present the possibility to achieve energy-field-induced energy tuning for the entangled-photon emission from QDs through simultaneous application of the uniaxial stress and the vertical electric field. 

In the absence of in-plane magnetic fields, the coupling between the dark exciton and the bright exciton states is negligible. When the uniaxial stress, $p$, and the vertical electric field are applied to QDs in our device, the FSS ($s$) and the exciton polarization angle ($\theta$) (see Fig.~2a and 2b) are given by

\begin{eqnarray}
s = \sqrt{[\alpha p + 2(\delta(0)+\gamma_\text{d}/2\times F_\text{d})]^2+4(\kappa+\beta p)^2}\\
\text{tan}~\theta_\pm = \frac{-2(\delta(0)+\gamma_\text{d}/2\times F_\text{d})-\alpha p \mp s}{2(\kappa+\beta p)},
\end{eqnarray}   
where $\alpha$, $\beta$ are the external stress related parameters, $\kappa$ is related to the QD structure asymmetry and $\delta(0)$ accounts for the mesoscopic structure of the QD at $F_\text{p} = 0$ kV cm$^{-1}$, $\gamma_\text{d}$ is proportional to the difference of the exciton dipole moments (see the derivation in Supplementary Note 1). 

From equation~(1), it is obvious that $s = 0$ can be theoretically obtained when $p$ and $F_\text{d}$ take the critical values: $p_\text{c} = -\kappa/\beta$ and $F_\text{d}^\text{c} = -[\alpha p_\text{c} + 2\delta(0)]/\gamma$. This suggests that the FSS can be universally tuned to zero through application of both the uniaxial stress and the vertical electric field in our device.~Now we consider a special case in which the following conditions are met: (1) the external stress $p$ is well aligned along the crystal axis [110] ([1-10]) of the GaAs and (2) the exciton polarization angle at $F_\text{p} =0 $ kV cm$^{-1}$, denoted by $\theta_0$, is precisely oriented along the [110] ([1-10]) axis of the GaAs. Subsequently,  $\beta$ = 0 and $\kappa$ = - $s_0 \times$ sin$(2\theta_0)/2$ = 0, where \textit{$s_0$} is the FSS at $F_\text{p} =0 $ kV cm$^{-1}$ (see Supplementary Note 1). In this context, equation (1) will reach the minimum value at zero when $F_\text{d}$ and \textit{p} take
\begin{equation}
p_\text{c} = - \frac{2\delta(0)+\gamma_\text{d} F_\text{d}^\text{c}}{\alpha}.
\end{equation}
Strikingly, equation~(3) implies that the critical stress $p_\text{c}$ is no longer restricted to a single value but shows a linear dependence on the externally applied $F_\text{d}$. This indicates that the critical uniaxial stress applied to eliminate the FSS of QDs can be linearly tuned by the vertical electric field. As a consequence, energy-tunable entangled-photon emission can be expected from our device. 
\begin{figure*}[t]
	\centering
	\includegraphics[scale=0.9]{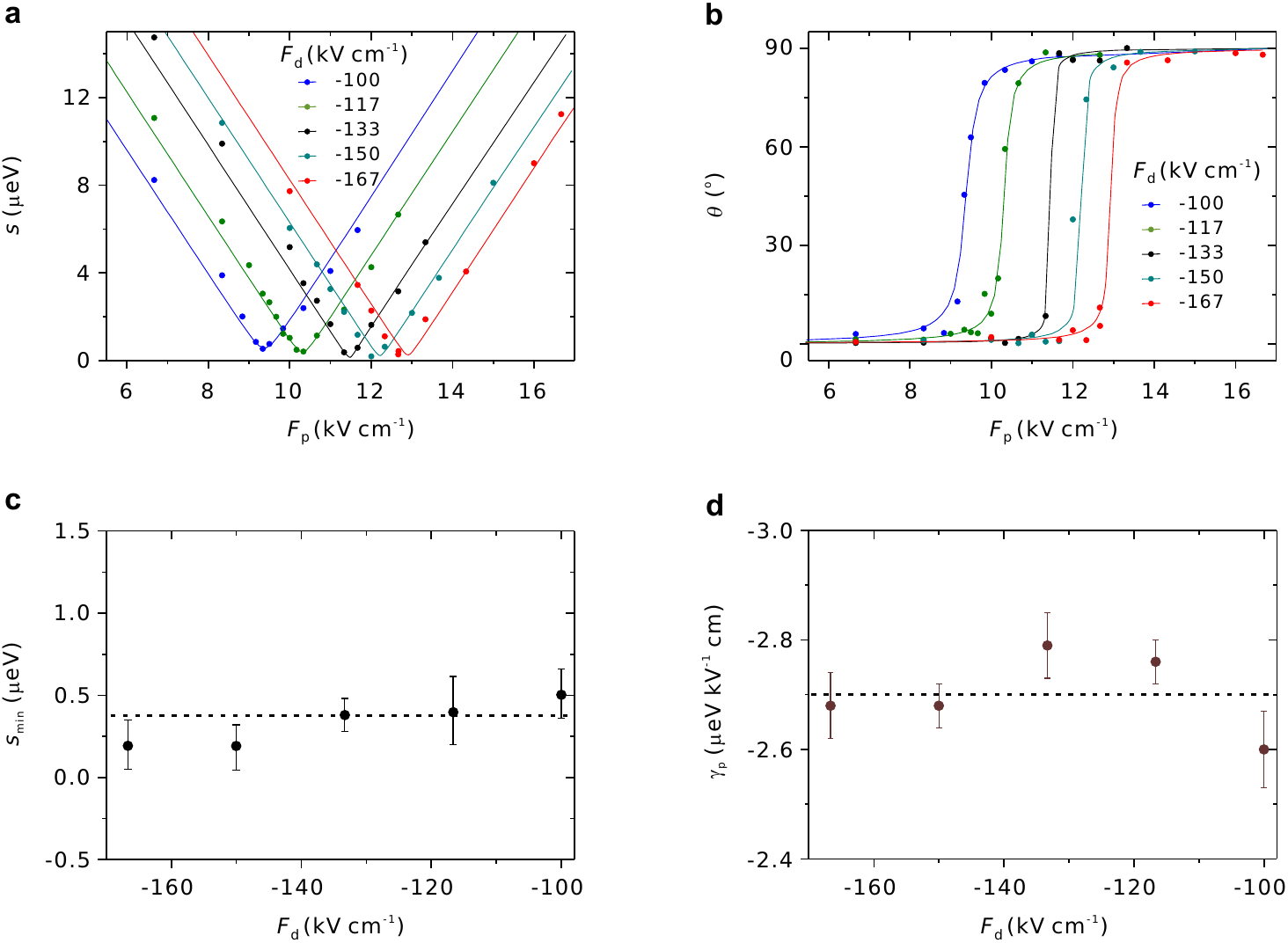}
	\caption{\textbf{$|$ Behaviors of $s$ and $\theta$ through simultaneous application of $V_\text{p}$ and $V_\text{d}$}. \textbf{a}, \textbf{b} Magnitude of $s$ and $\theta$ as a function of $F_\text{p}$ for five different values of $F_\text{d}$. Solid lines are theoretical fits using equations (1) and (2). \textbf{c}, The minimum FSS, $s_\text{min}$, tuned by anisotropic strain fields ($F_\text{p}$) for the five different $F_\text{d}$. \textbf{d}, Tuning rate of \textit{s} when the FSS is far away from the minimum value for the different applied $F_\text{d}$.}
\end{figure*}  

Experimentally, the above two conditions can be satisfied by carefully aligning the GaAs crystal axis [1-10] along the axis of the uniaxial stress, and selecting a QD whose exciton polarization angle $\theta_0$ is precisely aligned along the GaAs [110] ([1-10]) direction.~These two stringent requirements can be verified by monitoring the change of $s$ and $\theta$ as $p$ is solely applied to the QD. Fig.~2c and 2d show the magnitude of $s$ and $\theta$ as a function of $F_\text{p}$ at $V_\text{d} = 0 V$.~$\theta$ is found to be 180.23 $\pm$ 0.02$^{\circ}$ at $V_\text{d} = 0 V$, which indicates a precise alignment of the exciton polarization direction along the GaAs [1-10] direction (see Methods).~Moreover, as $F_\text{p}$ is swept from -3.3 to 10 kV~cm$^{-1}$, we find a drastic change in the FSS, whereas $\theta$ remains almost unchanged. It should be noted that such behaviors can occur only when both of the GaAs [110] ([1-10]) direction and the exciton polarization direction are well aligned along the uniaxial stress axis \cite{Zhang15}. Additionally, we investigate the change of $s$ and $\theta_\text{0}$ as a function of $F_\text{d}$ at $V_\text{p}$ = 0~V and the results are shown in Fig.~2e and 2f. Interestingly, a similar behavior for the application of $F_\text{p}$ is observed. The FSS is obviously reduced while $\theta_\text{0}$ stays almost constant at $\sim$180$^{\circ}$ as $F_\text{d}$ is varied. Such behavior further confirms the precise alignment of the exciton polarization direction along the GaAs optical axes \cite{Bennett10}.  

In addition, in both cases, the FSS exhibits a linear change in magnitude with electric fields $F_\text{p}$ and $F_\text{d}$, and the gradient is found to be $\gamma_\text{p}$ = -2.81 $\pm $ 0.05  and  $\gamma_\text{d}$ = -0.104 $\pm $ 0.007 $\mu$eV kV$^{-1}$ cm, respectively. We can clearly see that the tuning rate for $F_\text{p}$ is approximately 28 times larger than that of $F_\text{d}$. In contrast to this FSS change, the energy shift induced by $F_\text{p}$ and $F_\text{d}$ shows converse tuning trend. The energy shift due to $F_\text{d}$ is found to be 4.6 meV, which is much larger than the energy shift of 1.8 meV by $F_\text{p}$. From these direct observations, therefore, it is intuitive to choose $F_\text{d}$ as the energy tuning mechanism, and $F_\text{p}$ as the FSS tuning mechanism in our later studies.  
\begin{figure*}[t!]
	\centering
	\includegraphics[scale=0.9]{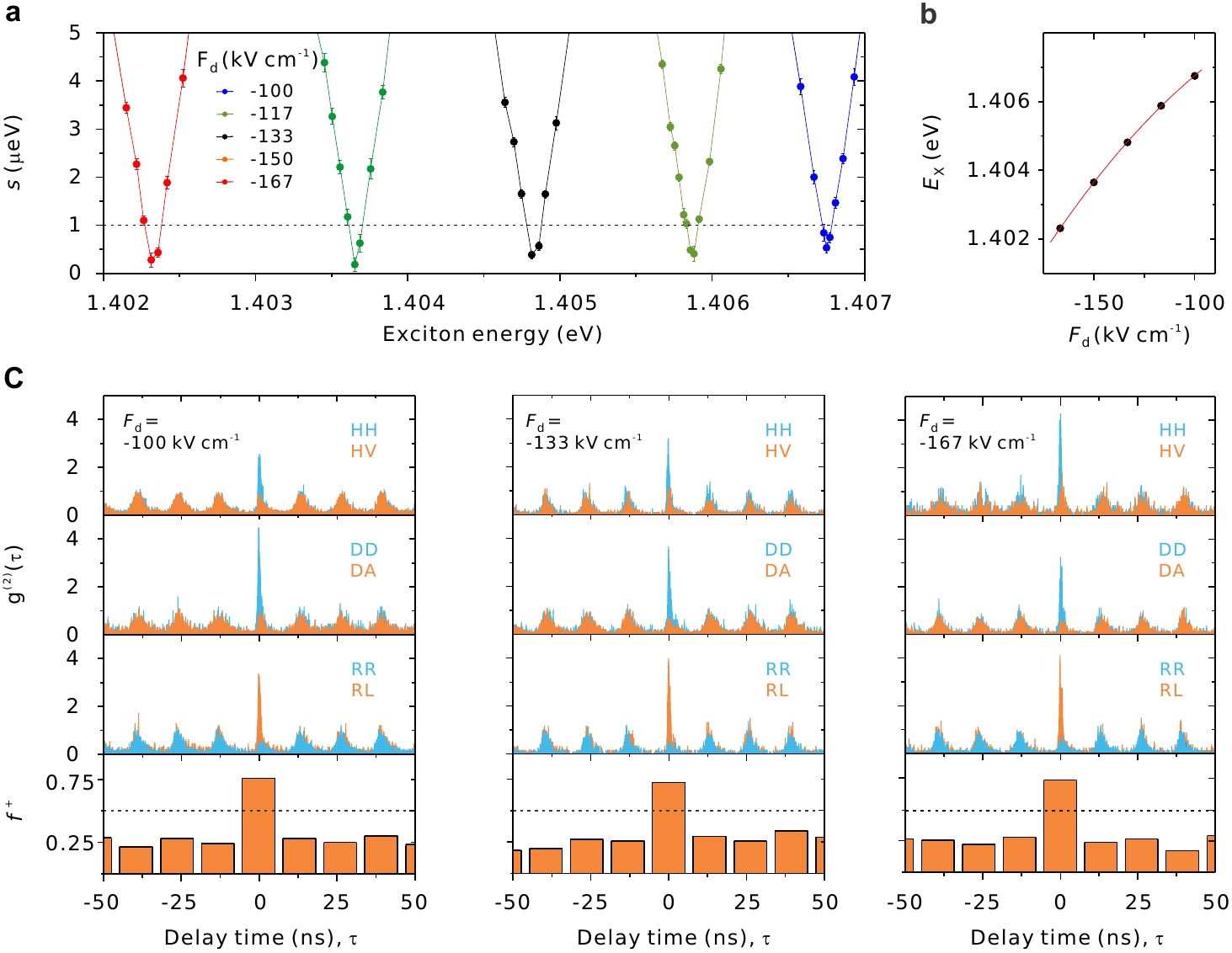}
	\caption{\textbf{$|$ Electric-field-induced energy tuning of triggered entangled-photon emission from a single QD}. \textbf{a}, Magnitude of $s$ as a function of $E_\text{X}$ for five different values of $F_\text{d}$ applied to the diode. \textbf{b}, E$_X$ when $s = s_\text{min}$ as a function of $F_\text{d}$, the solid line is theoretical fit. \textbf{c}, Normalized correlation functions for co- and cross-polarized XX and X photons in HV, DA and RL bases when $s=s_\text{min}$ for three $F_\text{d}$: -100, -133.3 and -167 kV cm$^{-1}$, respectively. The bottom panels show the entanglement fidelity to the Bell state $|\Psi^+>$. The dashed lines indicate the classical limit with the value of 0.5. }
\end{figure*}

In order to access the minimum FSS ($s_\text{min}$) of the QD, we increase $F_\text{p}$ to apply larger uniaxial stress to the QD. Fig. 3a shows the behavior of $s$ as a function of $F_\text{p}$ for five different values of $F_\text{d}$.~It is evident that the magnitude of $s$ at different $F_\text{d}$ decreases to an almost zero minimum value of $s_\text{min} \approx 0~\mu$eV as appropriate uniaxial stresses are applied. As already predicted in the theory (see Supplementary Note 1 and Supplementary Figure 1), the critical value of $F_\text{p}$ at $s_\text{min} \approx 0$ can be continuously shifted with $F_\text{d}$.~In addition, the behavior of $\theta$ for the five different values of $F_\text{d}$ are also investigated, as shown in Fig. 3b. When $F_\text{p}$ is increased, $\theta$ remains unchanged as $s$ is far away from its $s_\text{min}$, and then it shows a sharp change as $s$ approaches the minimum value. Further increase of $F_\text{p}$ leads to an inversion of $\theta$ from 0$^{\circ}$ to 90$^{\circ}$, which, remarkably, provides an experimental signature that $s_\text{min}$ can be experimentally tuned to zero regardless of the limited spectral resolution of the setup. 

Equations (1) and (2) are used to fit the experimental data as the solid lines show in Fig. 3a and 3b (see Supplementary Note 2 and Supplementary Table 1). From the theory, the minimum FSS value for each $F_\text{d}$, following the form $s_\text{min} = |s_0\times$ sin$(2\theta_\text{0})|$ \cite{Gong11}, is determined by the initial FSS $s_0$ and polarization angle $\theta_\text{0}$ at $F_\text{p}$ = 0 kV~cm$^{-1}$. The predicted values, before varying $F_\text{p}$, are found to be 0.44 $\pm$ 0.12, 0.13 $\pm$ 0.10, 0.26 $\pm$ 0.07, 0.20 $\pm$ 0.16 and 0.10 $\pm$ 0.14 $\mu$eV for different $F_\text{d}$ at -100, -117, -133, -150 and -167 kV~cm$^{-1}$, respectively. These values are found to be consistent with our experimentally observed values (see Fig. 3c). The above theoretical fits and well-predicted values of $s_\text{min}$ for different $F_\text{d}$ indicate an excellent agreement between our simple model and the experimental data. Even more, we extract the FSS tuning rate $\gamma_\text{p}$ over $F_\text{p}$ as a function of $F_\text{d}$ (see Fig. 3d). $\gamma_\text{p}$ remains constant at about -2.7  $\mu$eV kV$^{-1}$ cm, which reflects that the application of $F_\text{d}$ does not affect the tuning effects of $F_\text{p}$ but only changes the position of the critical stress for $s_\text{min} \approx 0 ~\mu$eV.

Instead of using $F_\text{p}$, Fig. 4a plots $s$ as a function of the exciton emission energy, $E_\text{X}$, for five different values of $F_\text{d}$. We can see that $s_\text{min}\approx 0~\mu$ eV can be achieved over a broad spectral range of about 5 meV. The value of $E_\text{X}$ as a function of $F_\text{d}$ at $s_\text{min} \approx 0 $ is shown in Fig. 4b.~This energy shift is well ascribed to the QCSE that can be formulated by $E = E_\text{X}^0 + \eta F_\text{d} + \chi F_\text{d}^2 $ \cite{Bennett10}, where $E_\text{X}^0$ is the energy in the absence of $F_\text{d}$, $\eta$ is the permanent dipole moment of the QD in the $z$ axis, $\chi$ is the polarizability. From the theoretical fit we can extract $\eta$ = -4.0 $\pm$ 0.4 $\mu$eV~cm~kV$^{-1}$ and $\chi$ = -0.26 $\pm$ 0.01 $\mu$eV~cm$^2$~kV$^{-2}$.   

The ability to eliminate the FSS of the QD and to tune its photon emission energy over a broad range allows us to demonstrate triggered energy-tunable polarization entangled-photon emission from our device. We carried out polarization-resolved co- and cross-correlations between the XX and X photons emitted by the QD when $s$ is tuned to $s_\text{min} \approx 0~\mu$eV for the five different values of $F_\text{d}$. Fig. 4c shows the polarization-correlation results when the energy of the X photon is tuned to $E_1$ = 1.4068 eV, $E_2$ = 1.4048 V and $E_3$ = 1.4023 eV, corresponding to $F_\text{d}$ = -100, -133 and -167 kV cm$^{-1}$, respectively. All of the correlation functions show periodic correlation peaks with a well-separated temporal distance of 13.1 ns, which is consistent with the repetition rate of 76 MHz of the excitation laser, suggesting a triggered generation of the X and XX photon pairs. As expected for the two-photon polarization entanglement, strong correlations are always observed in the rectilinear $\{$H, V$\}$ and diagonal $\{$D, A$\}$ basis, while anti-correlation is observed in the circular basis $\{$R, L$\}$ for co-polarized two photons (blue curves). In contrast, for cross-polarized two photons, we obtain anti-correlations in the $\{$H, V$\}$ and the $\{$D, A$\}$ bases but correlation in the $\{$R, L$\}$ basis (orange curves). In order to obtain the entanglement fidelity ($f^+$) to the Bell state $|\Psi^+\rangle$=$[|H_\text{XX}$$H_X\rangle$ + $|V_\text{XX}V_X\rangle$]$/\sqrt{2}$ \cite{Benson00},~we calculate the degrees of the polarization correlation, $C_\text{HV}$, $C_\text{DA}$ and $C_\text{RL}$ in the $\{$H, V$\}$, $\{$D, A$\}$ and $\{$R, L$\}$ basis respectively from the normalized co- and cross-polarization correlations, $g^{(2)}(\tau)$ (see Methods). By taking the coincidence counts integrated in a 1.5 ns temporal window centered at zero delay time, we measure $C_\text{HV}$= 0.49 $\pm$ 0.07,  $C_\text{DA}$ = 0.61 $\pm$ 0.07,  $C_\text{RL}$ = -0.79 $\pm$ 0.09 for $E_1$; $C_\text{HV}$= 0.55 $\pm$ 0.14,  $C_\text{DA}$ = 0.69 $\pm$ 0.06,  $C_\text{RL}$ = -0.71 $\pm$ 0.09 for $E_2$ and $C_\text{HV}$= 0.57 $\pm$ 0.04,  $C_\text{DA}$ = 0.68 $\pm$ 0.08,  $C_\text{RL}$ = 0.77 $\pm$ 0.04 for $E_3$. As a result, $f^+$ is found to be  $f^+_1$ = 0.72 $\pm$ 0.04, $f^+_2$ = 0.74 $\pm$ 0.03 and $f^+_3$ = 0.76 $\pm$ 0.03. They all exceed the classical limit of 0.5, thus proving triggered and energy-tunable generation of entangled-photon pairs from our device.

In summary, we have demonstrated a viable concept and a facile approach to realize electric-field-induced energy tuning of on-demand entangled-photon emission from a QD. This device consists of an electrically-tunable diode nanomembrane integrated onto the PMN-PT crystal capable of exerting uniaxial stress to QDs.~We have experimentally demonstrated electric-field-induced energy tuning of about 5 meV for entangled-photon emission from QDs. High entanglement-fidelities well above 0.72 have been obtained for all tuned energies under the applied vertical electric field, which indicates the entanglement of the QD emitter can be well-preserved in response to the external strain and electric fields. It is worth noticing that the energy tuning range is not limited to about 5 meV, and energy shift of about 25 meV is possible for this entangled-photon source when extending the vertical electric field to a range of $-300 <F_\text{d}<0 $ kV cm$^{-1}$ (see Supplementary Figure 2). Notably, the experimental realization of electric-field control over the energy of entangled-photon emission from QDs can alleviate the slow creep effect induced by the strain field. Therefore, the device demonstrated in this work has practical advantages over other schemes \cite{Trotta16,Chen16,Trotta15,Wang15}, thus constituting an important step towards realization of entanglement swapping, as well as other hybrid quantum systems for quantum network applications. 

In addition to the above newly revealed feature in such hybrid piezoelectric-semiconductor QD devices, we point out that application of the uniaxial stress and the vertical electric field cannot ``universally'' eliminate the FSS for every QD in the device.~Since the GaAs [1-10] ([110]) direction is precisely aligned along the uniaxial stress axis, we obtain $\beta\approx 0$ and $\alpha\neq 0$ from the theory. For the rest of QDs whose exciton polarization angles are not well oriented along the [1-10] ([110]) direction of the GaAs,  $\kappa$ = -$s_0 \times$ sin$(2\theta_\text{0})/2$ $\neq$ 0 as $\theta_\text{0} \neq$ 0$^{\circ}$ or 90$^{\circ}$.~Therefore, the critical stress required to cancel the FSS can be written as $p_\text{c} = -\kappa/\beta \rightarrow\infty$. This indicates that experimental realization of $s = 0$ for these dots requires an extremely large magnitude of the uniaxial stress. In order to make an universal FSS tuning possible, the orientation of the GaAs [110] ([1-10]) axis should be slightly rotated away from the uniaxial stress axis. We leave this for the future experimental study. 

\section{Methods}
\textbf{Sample and device fabrication}. The studied sample was grown on a (001) GaAs substrate by solid-source molecular beam epitaxy (MBE). It consists of a \textit{p-i-n} heterostructure diode with a 150-nm thick intrinsic GaAs/Al$_\text{0.4}$Ga$_\text{0.6}$As quantum well, in which low density of In(Ga)As QDs are embedded in the middle. The electric field $F_\text{d}$ applied on the diode is given by $F_\text{d}=-(V_\text{bi}-V_\text{d})/d$ (with $V_\text{bi}\sim-2.0~V$ being the diode built-in potential and $d$ = 150 nm the intrinsic region thickness).~The entire diode structure was grown on a 100 nm-thick Al$_\text{0.75}$Ga$_\text{0.25}$As sacrificial layer. In order to process the diode nanomembranes, standard UV photolithography and wet chemical etching were used to fabricate mesa structures with size of 120$\times$160 $\mu$m$^2$. The longer edge of the GaAs membrane was processed along [1-10] crystal axis of the GaAs and  was carefully aligned along the \textit{y} axis of the PMN-PT actuator. The bonded gold layer on the bottom formed a \textit{p}-contact, while the \textit{n}-type contact was formed by depositing a gold pad with size of 50$\times$50 $\mu$m$^2$ on the top of the nanomembrane.
\newline

\textbf{Optical measurements.} The PL was excited by a mode-locked Ti:Sapphire laser with a central wavelength of 840 nm and was then collected by a 50$\times$ microscope objective with a numerical aperture of 0.42. The microscope objective is placed on the top of the nanomembrane collecting the photon emission from the area close to the metal contact. By inserting a half-wave plate and a linear polarizer directly after the collection lens, polarization-resolved measurements were performed in order to obtain the FSS \textit{vs} $F_\text{p}$ (or $F_\text{d}$). The exciton polarization is determined by aligning the fast optical axis of the polarizer along the [1-10] direction of the GaAs nanomembrane. The PL was directed to a spectrometer with 750 mm focus length, and the spectrum is analyzed using a nitrogen cooled charge-coupled device. The FSS is determined with an accuracy of sub-$\mu$eV.
\newline

\textbf{Polarization-resolved photon correlation measurements.} A non-polarizing 50:50 beam splitter is placed directly after the collection objective in order to divide the optical paths between two spectrometers, which are used to detect X and XX separately. After each spectrometer, a Hanbury-Brown Twiss setup, consisting of a polarizing beam splitter and two high efficiency single-photon avalanche detectors, is placed. Half- and quarter-wave were used to select the proper polarization basis. The temporal resolution of the system is about 400 ps. The entanglement is quantified by measuring the degree of correlation C, which is defined by
\begin{equation}
C_\text{basis}=\frac{g_{\text{XX,X}}^{(2)}(\tau)-g_{XX,\hat{X}}^{(2)}(\tau)}{g_\text{{XX,X}}^{(2)}(\tau)+g_{XX,\hat{X}}^{(2)}(\tau)}, \nonumber
\end{equation}
where $g_\text{{XX,X}}^{(2)}(\tau)$ and $g_{XX,\hat{X}}^{(2)}(\tau)$ are normalized second-order time correlations for co-polarized and cross-polarized XX and X photons, respectively. The fidelity $f^+$ is calculated by using the formula:  $f^+ =(1+C_\text{HV}+C_\text{DA}-C_\text{RL})/4$.

~
\textbf{Acknowledgments} The work was supported financially by BMBF QuaHL-Rep (Contracts no. 01BQ1032 and 01BQ1034), Q.Com-H (16KIS0106) and the European Union Seventh Framework Programme 209 (FP7/2007-2013) under Grant Agreement No. 601126 210 (HANAS). J. X. Zhang was supported by China Scholarship Council (CSC, No. 2010601008). Analysis of QDs spectra was accomplished by using the XRSP3 software developed by A. Rastelli. The authors thank B. Eichler, R. Engelhard, P. Atkinson and S. Harazim for the technical support on device fabrication. 

\textbf{{Author contributions}}
The sample was grown by E. Z and the device was fabricated by J. X. Z. with help from B. H., Y. C., R. K., M. Z., S. B. The work was conceived by J. X. Z. and guided by F. D and O. G. S. The optical measurements were performed by J. X. Z. and the results are discussed by all authors. J. X. Z. wrote the manuscript with input from all the other authors. 

\textbf{{Additional information}} The authors declare no competing financial interests.~Supplementary information accompanies this paper on www.nature.com/naturecommunications.~Reprinting and permissions information is available online at http://npg.nature.com/reprintsandpermissions. Correspondence and requests for materials should be addressed to J. X. Z.

\newpage
\pagestyle{plain}
\section{}
\pagebreak
\setcounter{equation}{0}
\setcounter{figure}{0}
\setcounter{table}{0}
\setcounter{page}{1}
\makeatletter
\renewcommand{\theequation}{S\arabic{equation}}
\renewcommand{\bibnumfmt}[1]{[S#1]}
\renewcommand{\citenumfont}[1]{S#1}
\renewcommand{\figurename}{Supplemental Figure}
\renewcommand{\tablename}{Supplemental Table}
\newpage
\newpage

\section{SUPPLEMENTARY INFORMATION}
\subsection{Supplementary Note 1: The effective two-level exciton Hamiltonian under a uniaxial stress and a vertical electric field}
In our experiment, the used piezoactuator has pseudo-cubic cut directions [100], [0-11] and [011], denoted by \textit{x}, \textit{y}, \textit{z} axis respectively.~When it is poled along the \textit{z} axis, in-plane strain fields with normal components $\varepsilon_\text{xx}$ along the \textit{x} axis and $\varepsilon_\text{yy}$ along the \textit{y} axis with opposite sign can be transferred to the QDs nanomembrane.~According to the piezoelectric coefficients $d_\text{31}\sim+420~pC N^{-1}$ along the \textit{x} axis and $d_\text{32}\sim-1140~pC N^{-1}$ along the \textit{y} axis \cite{Hans}, the in-plane strain anisotropy is estimated to be $\varepsilon_\text{xx} \approx -0.37\varepsilon_\text{yy}$.~Axes of the anisotropic strain fields, $\varepsilon_\text{xx}$ and $\varepsilon_\text{yy}$, were carefully aligned along the GaAs crystal axis [110] and [1-10] respectively.~Owing to the zinc blende crystal structure of the GaAs, the applied anisotropic strain fields induce anisotropic stresses $p_\text{xx}$ and $p_\text{yy}$ in the [110] and [1-10] directions of the diode nanomembrane. Their magnitudes are quantified by \cite{Chuang95}
\begin{equation}
\begin{pmatrix}
p_\text{xx} \\
p_\text{yy} 
\end{pmatrix}= \begin{pmatrix}
C_\text{11} & C_\text{12}\\
C_\text{12} & C_\text{11}
\end{pmatrix} \begin{pmatrix}
\varepsilon_\text{xx} \\
\varepsilon_\text{yy}
\end{pmatrix},
\end{equation}
where $C$ is the elastic stiffness tensor of the GaAs and $C_\text{11}$ = 11.88$\times 10^{10} N m^{-2}$, $C_\text{12}$ = 5.38$\times 10^{10} N m^{-2}$ \cite{Adachi85}. This gives $p_\text{yy} \approx 10~p_\text{xx}$.~Consequently, our anisotropic stresses can be thought as uniaxial stress $p$ and its orientation is in parallel with the [1-10] crystal axis of the GaAs nanomembrane. As such uniaxial stress is applied to self-assembled In(Ga)As/GaAs quantum dots (QDs), the tuning behaviors of the fine structure splitting (FSS, $s$) and the polarization angle, $\theta$, can be well explained by employing the uniaxial stress dependent exciton Hamiltonian proposed by M. Gong et al. \cite{Gong11s,Zhang15s}. 

Now we consider the impact of the vertical electric field, $F_\text{d}$, on the fine structure splitting (FSS, $s$), and the exciton poalrization direction ($\theta$) in our device. In fact, the effect of a vertical electric field is equivalent to the uniaxial stress along the [110] direction according to the symmetry analysis \cite{Wang12s}.~As already demonstrated in experiment, application of the vertical electric field would induce a coherent coupling between the two bright exciton states, which subsequently results in the suppression of the FSS \cite{Bennett10s,Ghali12s}. By combining the vertical-electric-field dependent exciton two-level Hamiltonian developed by A. J. Bennett and co-workers \cite{Bennett10s}, we obtain the effective Hamiltonian in the space spanned by the two bright exciton states    
\begin{widetext}
	\begin{equation}
	H = \begin{pmatrix}
	E_0+\alpha_1p+\delta(0)+\gamma_\text{d}/2F_\text{d} & \kappa + \beta p \\
	\kappa + \beta p & E_0+\alpha_2p -\delta(0)-\gamma_\text{d}/2F_\text{d}  
	\end{pmatrix},
	\end{equation} 
\end{widetext}
where $E_0$ is the energy for the degenerated bright exciton states of the QD without broken structure symmetry; $\alpha_i$ ($i= 1,2$) and $\beta$ are parameters related to the external uniaxial stress, $\gamma_\text{d}$ is proportional to the difference of the exciton dipole moments \cite{Gong11s}, $\delta(0)$ and $\kappa$ are determined by the QD mesoscopic structural asymmetry. Diagonalization of the above Hamiltonian gives the forms of $s$ and $\theta$ as shown in equations (1) and (2) in the main text:
\begin{eqnarray}
	s = \sqrt{[\alpha p + 2(\delta(0)+\gamma_\text{d}/2\times F_\text{d})]^2+4(\kappa+\beta p)^2}\\
	\text{tan}~\theta_\pm = \frac{-2(\delta(0)+\gamma_\text{d}/2\times F_\text{d})-\alpha p \mp s}{2(\kappa+\beta p)}. 
\end{eqnarray} 
By using $\delta = \delta(0)+\gamma_\text{d}/2\times F_\text{d}$, we then obtain:
\begin{eqnarray}
	s = \sqrt{(\alpha p + 2\delta)^2+4(\kappa+\beta p)^2}\\
	\text{tan}~\theta_\pm = \frac{-2\delta-\alpha p \mp s}{2(\kappa+\beta p)}. 
\end{eqnarray} 
Therefore, in analogy with the analysis in ref.~\cite{Gong11s}, for stress $p$ aligned along with the GaAs [1-10] ([110]) direction, we have $\beta \approx 0$ according to symmetry analysis. In the meantime, the parameters $k$ and $\delta$ can be simply deduced by using the experimentally observed FSS $s_0$ and the exciton polarization angle $\theta_\text{0}$ at $V_\text{p} = 0$: $k=-s_0 \times \sin(2\theta_\text{0})/2$ and $\delta=s_0 \times \cos(2\theta_\text{0})/2$. Most importantly, provided the uniaxial stress is precisely aligned along the GaAs [1-10] ([110]) direction, $k$ determines the minimum FSS tuned by $p$, while $\delta$ determines the magnitude of the critical uniaxial stress at which the minimum FSS is obtained. Since our $\delta$ shows a linear dependence on the vertical electric field, this indicate that the minimum FSS can be shifted by the applied electric field, and consequently energy-tunable shifting of the FSS.  

\subsection{Supplementary Note 2: The theoretical fit procedure for $s$ and $\theta$}
The behaviors of $s$ and $\theta$ are fitted using equations (1) and (2) in the main text. There are five parameters to be determined in order to fit our experimental data: $\alpha=\alpha_1-\alpha_2$, $\beta$, $\kappa$, $\delta(0)$ and $\gamma_\text{d}$. It is noticeable that $\alpha$ and $\beta$ are external stress related parameters. In the analysis of M. Gong et al. \cite{Gong11s}, for stress along the [110] ([1-10]) direction, $\beta\approx0$, while for stress along [100] or [010] direction, $\alpha\approx0$. Considering our specific experimental arrangement, the approximately uniaxial stress is intentionally aligned along the [1-10] direction of the GaAs and therefore it gives $\beta\approx0$.
Furthermore, as presented in the Supplementary Note 1, $\kappa$ and $\delta(0)$ account for the QD structural asymmetry and they can be determined experimentally by the initial FSS ($s_0$) and the polarization angle ($\theta_\text{0}$) at $F_\text{p} = 0$ kV cm$^{-1}$. We calculate $\kappa$ and $\delta(0)$ from our simple model
\begin{eqnarray}
\kappa = - s_0\times \sin(2\theta_\text{0})/2 \\
\delta(0) = s_0\times \cos(2\theta_\text{0})/2-\gamma_\text{d}/2\times F_\text{d},
\end{eqnarray}
where $\gamma_\text{d}$ can be experimentally determined by $\gamma_\text{d} = \partial s/\partial F_\text{d} = -0.104 \pm 0.007$ $\mu$eV kV$^{-1}$cm (see Fig. 2e in the main text).

In the theoretical fit procedure, we can experimentally use $s_0$ and $\theta_\text{0}$ at different $F_\text{d}$ to calculate the initial values of $\delta(0)$ and $\kappa$ (see Supplementary Table 1). Thereafter, a least-squares approach is employed, with which the collapsed residual $\chi^2_\text{tot} = \sum_\text{i}^{5}\chi^2_\text{i}$ is minimized, $\chi^2_\text{i}$ being the residual for different applied $F_\text{d}$. As a result, we find a group of values: $\alpha$ = -2.837 $\mu$eV kV$^{-1}$, $\beta$ = 0.006 $\mu$eV kV$^{-1}$,  $\kappa$ = -0.209 $\mu$eV, $\delta(0)$ = 6.5 $\mu$eV, $\gamma_\text{d}$ = -0.145 $\mu$eV kV$^{-1}$ cm. With these values, all the experimental data of $s$ and $\theta$ shown in Fig. 3a and 3b can be well fitted, which reflects the effectiveness of our simple model developed here. Even more, we perform theoretical simulations to $s$ and $\theta$ with these values and the results can be seen in Supplementary Figure 1. Of great interest, we extend the range of $F_\text{d}$ from -300 to 0 kV cm$^{-1}$, and find the minimum FSS can be well tuned below 0.5 $\mu$eV. The critical $F_\text{p}^\text{c}$ remains linear dependence on $F_\text{d}$, which indicates that larger energy tunability can be achieved (see Supplementary Figure 2). For $\theta$, the sharp handedness has been obtained in the whole range of $F_\text{d}$.

\begin{table*}[t!]
	\centering
	\caption{The experimentally determined parameters for fitting $s$ and $\theta$}
	\begin{tabular}{l >{\centering}m{2cm} >{\centering}m{2cm} >{\centering}m{2cm} c}
		\hline\hline
		$F_\text{d}$ (kVcm$^{-1}$)	& $s_\text{0}$ ($\mu$eV) & $\theta_\text{0}$ ($^{\circ}$) & $\kappa$ ($\mu$eV) & $\delta(0)$ ($\mu$eV) \\
		\hline
		-100		& 25.03 $\pm$ 0.22 & 0.52 $\pm$ 0.26  & -0.22 $\pm$ 0.06 & 7.31 $\pm$ 0.11 \\	
		-117		& 28.93 $\pm$ 0.18 & 0.13 $\pm$ 0.20 & -0.06 $\pm$ 0.05 & 8.39 $\pm$ 0.09  \\		
		-133		& 32.52 $\pm$ 0.24 & 0.23 $\pm$ 0.25 & -0.13 $\pm$ 0.07 & 9.19 $\pm$ 0.12 \\	
		-150		& 33.42 $\pm$ 0.47 & 0.17 $\pm$ 0.28 & -0.10 $\pm$ 0.08 & 8.91 $\pm$ 0.23 \\	
		-167	    & 32.47 $\pm$ 0.25 & 0.08 $\pm$ 0.25 & -0.04 $\pm$ 0.07 & 7.56 $\pm$ 0.12  \\
		\hline\hline
	\end{tabular}
\end{table*}

\begin{figure*}[h!]
	\includegraphics[scale=0.8]{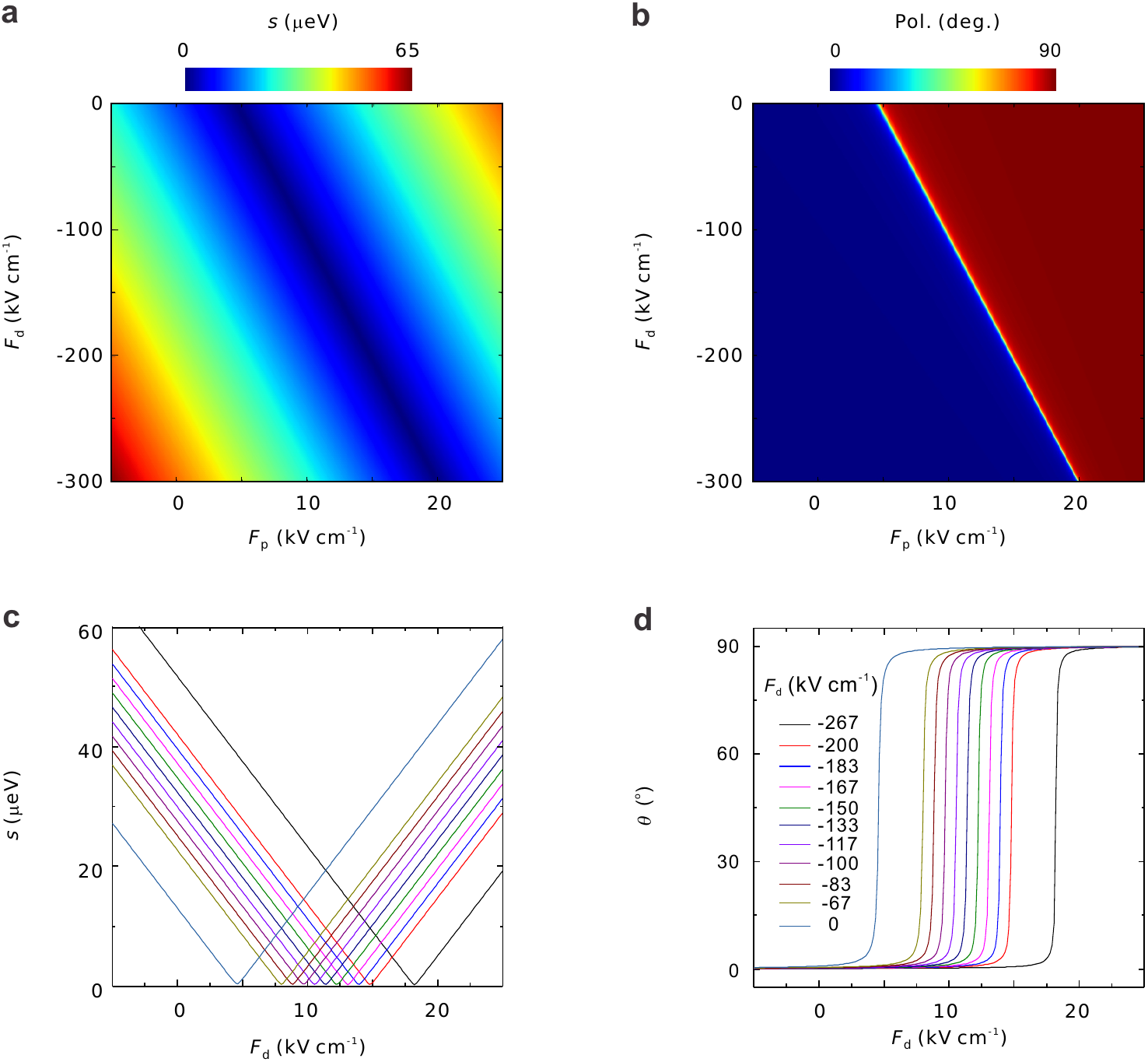}
	\caption{\textbf{Theoretical simulations of $s$ and $\theta$ under the uniaxial stress and the vertical electric field.} \textbf{a}, The behavior of $s$ as a function of $F_\text{d}$ and $F_\text{p}$ in the ranges of $-300 <F_\text{d}<0 $ kVcm$^{-1}$ and $-5 <F_\text{p}<25 $ kVcm$^{-1}$. In the whole range of $F_\text{d}$, application of $F_\text{p}$ always enables tuning of the FSS from a maximum value up to 60 $\mu$eV to almost zero value. Equivalently, at a fixed $F_\text{p}$, applying $F_\text{d}$ can also eliminate the FSS. In both cases, the critical electric fields for $s\approx 0$ show linear dependence, indicative of the energy shift for the QD emission. \textbf{b}, The magnitude of $\theta$ of the bright exciton emission. Sharp handedness in a range of (0$^{\circ}$, 90$^{\circ}$) is clearly seen when $s$ is tuned close to the minimum value for both $F_\text{d}$ and $F_\text{p}$. \textbf{c} - \textbf{d}, Behaviors of $s$ and $\theta$ extracted from \textbf{a} and \textbf{b} respectively. For different $F_\text{d}$, the FSS shows drastic change with a tiny lower bound and the polarization angle has sharp jump as $F_\text{p}$ is swept from $-5 <F_\text{p}<25 $ kVcm$^{-1}$.}
\end{figure*}

\begin{figure*}[h!]
	\centering
	\includegraphics[scale=1.0]{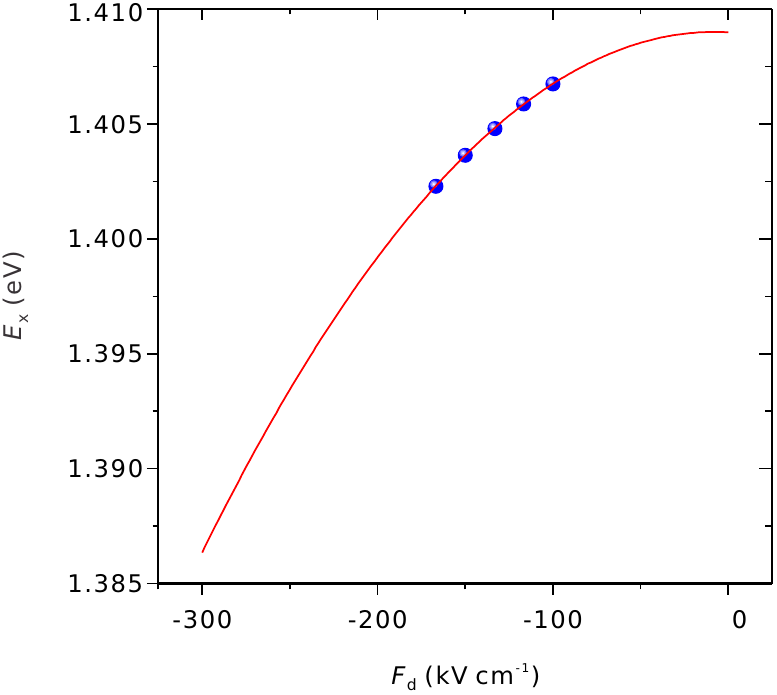}
	\caption{\textbf{Emission energy of the exciton photon at the minimum tuned FSS ($s_\text{min} \approx 0$) as a function of $F_\text{d}$}. The solid line is theoretical fit, from which we can find that the tuning range is quadratically increased by considering $F_\text{d}$ larger than those applied in the present study. Approximately 25 meV shift for the entangled-photon emission can be achieved when $F_\text{d}$ in the range of $-300 <F_\text{d}<0 $ kVcm$^{-1}$ is applied.}
\end{figure*}


\begin{thebibliography}{99}

		\bibitem{Gisin02}
		Gisin, N., Ribordy, G., Tittel, W. and Zbinden, H. Quantum cryptography. \textit{Rev. Mod. Phys.} \textbf{74}, 145 (2002).
	
		\bibitem{Bouwmeester97}
		Bouwmeester, D., et al. A. Experimental quantum teleportation. \textit{Nature} \textbf{390}, 575-579 (1997).
		
		\bibitem{Knill01}
		Knill, E., Laflamm R. and Milburn, G. J. A scheme for efficient quantum computation with linear optics. \textit{Nature} \textbf{409}, 46-52 (2001)
		
		\bibitem{Kok07}
		Kok, P. et al. Linear optical quantum computing with photonic qubits. \textit{Rev. Mod. Phys.} \textbf{79}, 135-174 (2007).
		
		\bibitem{Kimble08}
		Kimble, H. J. The quantum internet. \textit{Nature} \textbf{453}, 1023-1030 (2008). 
		
		\bibitem{Riedmatten04}
		de Riedmatten, H., Marcikic, I., Tittel, W., Zbinden, H., Collins, D. and Gisin, N.. Long distance quantum teleportation in a quantum relay configuration. \textit{Phys. Rev. Lett.} \textbf{92}, 047904 (2004).
		
		\bibitem{Pan98}
		Pan, J.-W., Bouwmeester, D., Weinfurter, H. and Zeilinger, A. Experimental entanglement swapping: entangling photons that never interacted. \textit{Phys. Rev. Lett.} \textbf{80}, 3891-3894 (1998).
		
		\bibitem{Riedmatten05}
		de Riedmatten, H. et al. Long-distance entanglement swapping with photons from separated sources. \textit{Phys. Rev. A} \textit{71}, 05302 (2005).
		
		\bibitem{Scarani04}
		Scarani, V., Ac\'{i}n, A., Ribordy, G. and  Gisin. N. Quantum Cryptography Protocols Robust against Photon Number Splitting Attacks for Weak Laser Pulse Implementations. \textit{Phys. Rev. Lett.} \textit{92}, 057901 (2004).
		
		\bibitem{Douse10}
		Dousse, A. et al. Ultrabright source of entangled photon pairs. \textit{Nature} \textbf{466}, 217-220 (2010).
		
		\bibitem{Versteegh14}
		Versteegh, Marijn A. M. et al. Observation of strongly entangled photon pairs from a nanowire quantum dot. \textit{Nat. Commun.}	\textbf{5}, 5298 (2014).
		
		\bibitem{Muller14}
		M\"{u}ller, M., Bounouar, S., J\"{o}ns, K. D.,	Gl\"{a}ssl, M. and Michler, P. On-demand generation of indistinguishable polarization-entangled photon pairs. \textit{Nat. Photon.} \textbf{8}, 224 - 228 (2014).
		
		\bibitem{Benson00}
		Benson, O., Santori, C., Pelton, M. and Yamamoto, Y. Regulated and Entangled Photons from a Single Quantum Dot, \textit{Phys. Rev. Lett.} \textbf{84}, 2513 (2000).
				
		\bibitem{Salter10}
		Salter, C. L., Stevenson, R. M., Farrer, I.,  Nicoll, C. A., Ritchie, D. A.	and Shields, A. J. An entangled-light-emitting diode, \textit{Nature} \textbf{465}, 594-597 (2010).
				
		\bibitem{Zhang15}
		Zhang, J. X. et al. High yield and ultrafast sources of electrically triggered entangled-photon pairs based on strain-tunable quantum dots. \textit{Nat. Commun.} \textbf{6}. 10067 (2015).
		
		\bibitem{Gammon96}
		Gammon, D., et al. Fine Structure Splitting in the Optical Spectra of Single GaAs Quantum Dots. \textit{Phys. Rev. Lett.} \textbf{76}, 3005 (1996).
		
		\bibitem{Bayer02}
		Bayer, M. et al. Fine structure of neutral and charged excitons in self-assembled In(Ga)As/(Al)GaAs quantum dots. \textit{Phys. Rev. B} \textbf{65}, 195315 (2002).		
		
		\bibitem{Stevenson05}
		Stevenson, R. M. et al. A semiconductor source of triggered entangled photon pairs. \textit{Nature} \textbf{439}, 179-182 (2005).
		
		\bibitem{Muller09}
		Muller, A. Fang, W. Lawal, J. and  Solomon, G. S. Creating Polarization-Entangled Photon Pairs from a Semiconductor Quantum Dot Using the Optical Stark Effect. \textit{Phys. Rev. Lett.} \textit{103}, 217402 (2009).

		\bibitem{Trotta16}
		Trotta, R. et al. Wavelength-tunable sources of entangled photons interfaced with atomic vapours. \textit{Nat. Commun.} \textbf{7}, 10375 (2016).

		\bibitem{Chen16}
		Chen, Y. et al. Wavelength-tunable entangled photons from silicon-integrated III–V quantum dots. \textit{Nat. Commun.} \textbf{7}, 10387 (2016).		
		
		\bibitem{Flagg10}
		Flagg, E. B. et al. Interference of Single Photons from Two Separate Semiconductor Quantum Dots. \textit{Phys. Rev. Lett.} \textbf{104}, 137401 (2010).	
		
	    \bibitem{Gold2014}
		Gold, P. et al. Two-photon interference from remote quantum dots with inhomogeneously broadened linewidths. \textit{Phys. Rev. B} \textbf{89}, 035313 (2014).	
		
		\bibitem{Giesz15}
		Giesz, V. et al. Cavity-enhanced two-photon interference using remote quantum dot sources. \textit{Phys. Rev. B} \textbf{92}, 161302(R) (2015).
		
		\bibitem{Patel2010}
		Patel, R. B.  et al. Quantum interference of electrically generated single photons from a quantum dot. \textit{Nanotechnology} \textbf{21}, 274011 (2010).		
			
		\bibitem{Zander09}
		Zander, T. et al. Epitaxial quantum dots in stretchable optical microcavities. \textit{Optics Express} \textbf{17}, 22452 (2009).
		
		\bibitem{Trotta12}
		Trotta, R. et al. Nanomembrane quantum-light-emitting diodes integrated onto piezoelectric actuators. \textit{Adv. Mater.} \textbf{24}, 2668–2672 (2012).
		
		\bibitem{Rastelli12}
		Rastelli, A. et al. Controlling quantum dot emission by integration of semiconductor nanomembranes onto piezoelectric actuators. \textit{Physica Status Solidi (b)} \textbf{249}, 687-696 (2012).
		
		\bibitem{Bennett10}
		Bennett, A. J. et al. Electric-field-induced coherent coupling of the exciton states in a single quantum dot. \textit{Nat. Phys.} \textbf{6}, 947 - 950 (2010).

		\bibitem{Zhang13}
        Zhang, J. X. et al. A nanomembrane-based wavelength-tunable high-speed single-photon-emitting diode. \textit{Nano Lett.} \textbf{13}, 5808–5813 (2013).
        
        \bibitem{Trotta121}
        Trotta, R. et al. Universal Recovery of the Energy-Level Degeneracy of Bright Excitons in InGaAs Quantum Dots without a Structure Symmetry. \textit{Phys. Rev. Lett.} \textbf{109}, 147401 (2012).
        		
        \bibitem{Trotta14}
        Trotta, R. et al. Highly entangled photons from hybrid piezoelectric-semiconductor quantum dot devices. \textit{Nano Lett.} \textbf{14}, 3439–3444 (2014).
        
        \bibitem{Plumhof11}
        Plumhof, J. D. et al. Strain-induced anticrossing of bright exciton levels in single self-assembled GaAs/Al$_x$Ga$_\text{1− x}$As and In$_x$Ga$_\text{1−x}$As/GaAs quantum dots. \textit{Phys. Rev. B} \textbf{83}, 121302 (2011).
        
        \bibitem{Zhang16}
        Zhang, J. X., Huo, Y. H., Ding, F. and Schmidt, O. G. Energy-tunable single-photon light-emitting diode by strain fields. \textit{Appl. Phys. B} \textbf{122}, 1–7 (2016).
		
		\bibitem{Gong11}
		Gong, M., Zhang, W., Guo, G.-C. and He, L. X. Exciton Polarization, Fine-Structure Splitting, and the Asymmetry of Quantum Dots under Uniaxial Stress, \textit{Phys. Rev. Lett.} \textbf{106}. 227401 (2011).
		
		\bibitem{Trotta15}
		Trotta, R. et al. Energy-Tunable Sources of Entangled Photons: A Viable Concept for Solid-State-Based Quantum Relays. \textit{Phys. Rev. Lett.} \textbf{114}, 150502 (2015).
		
		\bibitem{Wang15}
		Wang, J., Gong, M., Guo, G.-C. and He, L. Towards scalable entangled photon sources with self-assembled InAs/GaAs quantum dots. \textit{Phys. Rev. Lett.} \textbf{115}, 067401 (2015)

		
\end{thebibliography}

\begin{thebibliography}{99}
	\bibitem{Hans}
	Han, P., Yan, W., Tian, J., Huang, X. and Pan, H. Cut directions for the optimization of piezoelectric coefficients of lead magnesium niobate-lead titanate ferroelectric crystals. \textit{Appl. Phys. Lett.} \textbf{86}, 052902 (2005).
	
	\bibitem{Chuang95}
	Chuang, S. L. \textit{Physics of optoelectronic devices}. (wiley New York, 1995).
	
	\bibitem{Adachi85}
	Adachi, S. GaAs, AlAs, and Al$_x$Ga$_\text{1- x}$As: Material parameters for use in research and device applications. \textit{Journal of Applied Physics} \textbf{58}. R1 - R29 (1985)
	
	\bibitem{Gong11s}
	Gong, M., Zhang, W., Guo, G.-C. and He, L. X. Exciton Polarization, Fine-Structure Splitting, and the Asymmetry of Quantum Dots under Uniaxial Stress. \textit{Phys. Rev. Lett.} \textbf{106}. 227401 (2011).
	
	\bibitem{Zhang15s}
	Zhang, J. X. et al. High yield and ultrafast sources of electrically triggered entangled-photon pairs based on strain-tunable quantum dots. \textit{Nat. Commun.} \textbf{6}. 10067 (2015).
	
	\bibitem{Wang12s}
	Wang, J. P., Gong, M., Guo, G-C. and He, L. X. Eliminating the fine structure splitting of excitons in self-assembled InAs/GaAs quantum dots via combined stresses. \textit{Appl. Phys. Lett.} \textbf{101}, 063114 (2012).
			
	\bibitem{Bennett10s}
	Bennett, A. J. et al. Electric-field-induced coherent coupling of the exciton states in a single quantum dot. \textit{Nat. Phys.} \textbf{6}, 947 - 950 (2010).
	
	\bibitem{Ghali12s}
	Ghali, M., Ohtani, K., Ohno, Y. \& Ohno, H. Generation and control of polarization-entangled photons from GaAs island quantum dots by an electric field. \textit{Nature Commun.} \textbf{3}, 661 (2012).
\end{thebibliography}
\end{document}